# *Revisiting the Abraham-Minkowski Dilemma*

*Mário G. Silveirinha*[*]

*(1)University of Lisbon–Instituto Superior Técnico and Instituto de Telecomunicações, Avenida Rovisco Pais, 1, 1049-001 Lisboa, Portugal*

**Abstract**

Here, the Abraham-Minkowski controversy on the correct definition of the light momentum in a macroscopic medium is revisited with the purpose to highlight that an effective medium formalism necessarily restricts the available information on the internal state of a system, and that this is ultimately the reason why the dilemma has no universal solution. Despite these difficulties, it is demonstrated that in the limit of no material absorption and under steady-state conditions, the time-averaged light (kinetic) momentum may be unambiguously determined by the Abraham result, both for bodies at rest and for circulatory flows of matter. The implications of these findings are discussed in the context of quantum optics of moving media, and we examine in detail the fundamental role of the Minkowski momentum in such a context.

---

[*] To whom correspondence should be addressed: E-mail: *mario.silveirinha@co.it.pt*

# I. Introduction

The definition of the electromagnetic field momentum in a macroscopic material body is a fascinating problem, and is perhaps the longest standing quandary in classical electrodynamics [1, 2]. There are two main rival theories for the momentum of light, which were introduced more than 100 years ago by Minkowski and Abraham [1, 2]. Compelling and elegant *gedanken* experiments related to the center of mass energy favor Abraham's result [3, 4]. Other powerful arguments related to light diffraction and to Heisenberg's uncertainty principle appear to favor Minkowski's momentum [5]. A few now classical experiments attempted to settle the dispute (e.g., [6]-[8]) and their results at first sight suggest that Minkowski's momentum gives the correct answer. However, careful consideration of all the forces acting on the relevant material bodies shows that the experimental results are fully compatible with Abraham's momentum [9-10]. A seminal experiment was performed by Walker [1, 11] and verified the so-called Abraham component of the Lorentz force, thereby providing rather strong experimental evidence in favor of Abraham's theory.

The current status of the problem is a little hazy [12]. The most influential articles in recent years are perhaps those of Pfeifer *et al* [2] and of Barnett [13]. The point of view of Pfeifer *et al* is that the problem is now solved and that a formulation based on Abraham's momentum is fully equivalent to a formulation relying on Minkowski's momentum, being the adopted approach a matter of personal choice. The central argument is that the division of the total energy-momentum tensor into light and matter components is entirely arbitrary [4]. In fact, provided the *total* energy-momentum tensor is the same for two different frameworks (e.g., as in Eqs. (40)-(43) of Ref. [2]), then the optical (Lorentz) force acting on a given macroscopic material element is independent of the adopted light momentum formula, and hence it is impossible to favor any specific theory. Even though we agree with the general conclusions of Ref. [2], we would like to make the following points. The first point is that Ref. [2] does not really solve the dilemma, but rather highlights that the Minkowski and Abraham forms of the light momentum can be reconciled by choosing suitable material parts of the energy-momentum tensor that guarantee that the dynamics of a light-matter system is effectively independent of the formalism. Specifically, in Ref. [2] the material part of the energy-momentum tensor is chosen in such a way that the momentum of matter is given by the standard kinetic momentum ($\mathbf{p}_{kin}$) in Abraham's formalism, whereas in Minkowski's formalism it corresponds to a canonical momentum ($\mathbf{p}_{can}$). In our perspective, the most relevant question is not if the Minkowski and Abraham forms can be made equivalent with a judicious choice of the material part of the energy-momentum tensor. The important question is how to write the total momentum ($\mathbf{p}_{tot}$) of a given material element in terms of the kinetic momentum and of the instantaneous *macroscopic* electromagnetic fields. The answer provided by Eqs. (40)-(43) of Ref. [2] is $\mathbf{p}_{tot} = \mathbf{p}_{kin} + \mathbf{p}_{EM}$ with $\mathbf{p}_{EM} = \frac{1}{c^2} \int_{material} \mathbf{E} \times \mathbf{H}\, dV$, independent of the adopted (Abraham or Minkowski) formalism. This leads to the second point that we wish to highlight, which is that the second term ($\mathbf{p}_{EM}$) of the total momentum, let us call it kinetic momentum of light, is not necessarily settled, and may be challenged by future experiments because $\mathbf{p}_{kin}$ can be unambiguously measured and the total momentum transferred to a material body can also be determined, as further discussed later.

The theory of Barnett in Ref. [13] agrees on its essence with Pfeifer *et al* and he also emphasizes that the total momentum has different decompositions. In particular, Barnett underscores that the Abraham and Minkowski momenta correspond to kinetic and canonical light momenta, respectively, and thus are both physically meaningful. Indeed, we wish to underline that the kinetic and canonical momenta of light are both equally significant and fundamental. Another nice contribution by Barnett is that he explains why the momentum of a light quantum (photon) in a material is given by the Minkowski momentum, drawing attention to the fact that light quanta in dielectrics are quasi-particles resulting from the hybridization of the light and matter degrees of freedom [14]. In other words, in a material a light quantum has a matter component [14].

A point that we wish to bring into discussion, and which connects strongly with this article's perspective, is that the difficulty in the determination of the (kinetic) momentum of light ($\mathbf{p}_{EM}$) is rather fundamental and results mainly from limitations inherent to macroscopic electrodynamics [1, 2, 4]. In the words of Brevik [1], the "problem has no unique solution", or quoting Pfeifer *et al* the difficulty in writing the total momentum in terms of the macroscopic fields is due to "our (incomplete) understanding of material science". The key idea that guides the present research is that expressing either the kinetic or canonical light momenta in terms of the macroscopic fields remains an open problem. This difficulty is at the heart of the Abraham-Minkowski controversy, which has thereby its origin in the incomplete picture provided by an effective medium formalism. The central question under analysis in this article is how to write $\mathbf{p}_{EM}$ in terms of the macroscopic fields.

We find that despite the above discussed difficulties, under steady-state conditions and for dispersive media with negligible absorption loss the standard theory of macroscopic electrodynamics may provide enough information about the microscopic degrees of freedom to enable us to unambiguously determine the kinetic momentum of light ($\mathbf{p}_{EM}$) in a material body. It is shown that in a transient regime things are not as straightforward due to an incomplete knowledge of the microscopic state of the system, and generally it may not be possible to exactly determine the light momentum without further information about the internal degrees of freedom of the material. Finally, we discuss the formula for the light momentum in systems with moving parts, and the implications of our findings in the context of quantum optics.

Our analysis is based on simple energy and momentum conservation laws and on the assumption that the effective medium theory is sufficiently accurate so that outside the relevant materials the macroscopic and microscopic fields are coincident. Thus, we do not make any *a priori* assumptions about the microstructure of the materials, constitutive relations, or about the formulas of the light momenta and Lorentz force in the medium when written in terms of the macroscopic fields. In these rather general conditions, it is found that the macroscopic kinetic light momentum in steady-state conditions may be unequivocally determined when the materials are non-dissipative. It should be noted that the use of conservation laws in the context of Abraham-Minkowski dilemma has a long history (see Ref. [2] and the references therein).

# II. The Abraham-Minkowski controversy

## A. *The dilemma*

The Abraham-Minkowski controversy is a century old dilemma on the correct form of the electromagnetic momentum in a macroscopic medium. Abraham proposed that the macroscopic momentum density is given by

$$\mathbf{g}_{EM}^{Ab} = \frac{1}{c^2}\mathbf{E}\times\mathbf{H}, \tag{1a}$$

whereas Minkowski defended the alternative form

$$\mathbf{g}_{EM}^{Mi} = \mathbf{D}\times\mathbf{B}. \tag{1b}$$

In modern literature it is well understood that the nature of the Abraham and Minkowski momenta is different, and the former is usually regarded as the kinetic light momentum while the latter is regarded as a canonical light momentum [2, 13].

The light momentum appears in classical electromagnetism in connection with the stress-tensor theorem, which establishes that the Lorentz force acting on a set of (microscopic) charged particles enclosed by some volume $V$ and surrounded by air is given by [15]:

$$\mathbf{F}_L^{mic} = -\frac{d}{dt}\mathbf{p}_{EM}^{mic} + \int_{\partial V}\hat{\mathbf{n}}\cdot\overline{\mathbf{T}}ds. \tag{2}$$

Here, $\partial V$ represents the boundary surface, $\overline{\mathbf{T}}$ is the microscopic stress-tensor, $\overline{\mathbf{T}} = \varepsilon_0\mathbf{e}\otimes\mathbf{e} + \mu_0^{-1}\mathbf{b}\otimes\mathbf{b} - \left(\frac{1}{2}\varepsilon_0\mathbf{e}\cdot\mathbf{e} + \frac{1}{2}\mu_0^{-1}\mathbf{b}\cdot\mathbf{b}\right)\underline{\mathbf{1}}$, and $\mathbf{p}_{EM}^{mic}$ is the (kinetic) momentum of the fields in the volume $V$:

$$\mathbf{p}_{EM}^{mic} = \int_V \varepsilon_0\mathbf{e}\times\mathbf{b}dV. \tag{3}$$

The microscopic (macroscopic) fields are denoted with lower (upper) case letters. The conundrum is that in an effective medium framework, when the material is regarded as a continuum, it is not obvious how to write $\mathbf{p}_{EM}^{mic}$ in terms of the macroscopic fields. The solution of this problem is the central objective of this work.

Fortunately, there are no similar difficulties in calculating the second term of the Lorentz force ( $\int_{\partial V} \hat{\mathbf{n}} \cdot \overline{\mathbf{T}} ds$ ). Indeed, the stress tensor $\overline{\mathbf{T}}$ is evaluated in the air region, and in this region the microscopic fields may be assumed identical to the macroscopic fields. In particular, one sees that in a stationary regime (e.g., for any periodic in time excitation) one has:

$$\left\langle \mathbf{F}_L^{mic} \right\rangle_T = \int_{\partial V} \hat{\mathbf{n}} \cdot \left\langle \overline{\mathbf{T}} \right\rangle_T ds . \tag{4}$$

where $\left\langle ... \right\rangle_T$ represents time-averaging in one period of oscillation. Thus, the time-averaged Lorentz force is independent of the electromagnetic momentum: it is simply determined by the stress-tensor at the material boundary, which has an unequivocal form in the air region, even in the framework of macroscopic electromagnetism. In fact, the Abraham-Minkowski dilemma has relevant physical consequences mainly in the context of transient phenomena [9, 11], and this explains some of the difficulties in settling the dispute experimentally in the early years. Indeed, $\mathbf{p}_{EM}^{mic}$ determines the *time dynamics* of the Lorentz force. Without knowing $\mathbf{p}_{EM}^{mic}$ it is impossible to predict exactly the motion of a polarizable material body in a transient regime.

The point of view of this article is that the macroscopic kinetic electromagnetic momentum density $\mathbf{g}_{EM}$ should ideally be some function of the macroscopic fields that when integrated over some generic region $V$ of space gives a momentum that agrees with the microscopic theory:

$$\mathbf{p}_{EM} \equiv \int_V \mathbf{g}_{EM} dV = \int_V \varepsilon_0 \mathbf{e} \times \mathbf{b} \, dV . \tag{5}$$

Thus, when $\mathbf{p}_{EM}$ is combined with the kinetic momentum of the material particles one obtains the total momentum inside $V$.

## B. Decompositions of energy and momentum

We are interested in the interactions between polarizable material bodies and the electromagnetic field. Since an exact relativistic treatment seems to be a hopeless task in the general context of macroscopic electrodynamics, in our theory any relativistic corrections in the formulas of physical quantities, such as energy, momentum, etc, on the order of $v_\alpha^2 / c^2$ are neglected. Here, $v_\alpha$ is a generic velocity, for example the velocity of a generic microscopic charged particle.

Within a $o(v_\alpha^2 / c^2)$ approximation ($o(...)$ represents a quantity with magnitude on the order of the term in brackets), the total kinetic momentum of a set of charged particles (with rest mass $m_\alpha$ and velocity $\mathbf{v}_\alpha$) is given by $\mathbf{p}_{kin}^{mic} = \sum_\alpha m_\alpha \mathbf{v}_\alpha = M\mathbf{v}$, where $M$ is the total mass and $\mathbf{v}$ is the center of mass velocity. The total momentum in a generic volume $V$ can be written as the sum of the electromagnetic momentum and the kinetic momentum (here it is supposed for simplicity that all the particles are associated with the same material body) [2]:

$$\mathbf{p}_{tot}^{mic} = \mathbf{p}_{EM}^{mic} + \mathbf{p}_{kin}^{mic} = \mathbf{p}_{EM}^{mic} + M\mathbf{v} . \tag{6}$$

The time derivative of the kinetic momentum is determined by Newton's law

$$\frac{d\mathbf{p}_{\text{kin}}^{\text{mic}}}{dt} = M\frac{d\mathbf{v}}{dt} = \mathbf{F}_{\text{L}}^{\text{mic}} + \mathbf{F}_{\text{ext}}^{\text{mic}}, \tag{7}$$

where $\mathbf{F}_{\text{ext}}^{\text{mic}}$ represents the sum of possible external forces (e.g., mechanical forces) acting on the body, and $\mathbf{F}_{\text{L}}^{\text{mic}}$ is Lorentz's force [Eq. (2)].

On the other hand, the total energy in volume $V$ is the sum of the electromagnetic, the kinetic, and the rest mass energies, $\tilde{E}_{\text{tot}}^{\text{mic}} = E_{\text{EM}}^{\text{mic}} + E_{\text{kin}}^{\text{mic}} + Mc^2$. The rest mass energy is included for a matter of consistency. The kinetic energy can be conveniently decomposed as the center of mass energy ($\frac{1}{2}Mv^2$) and a vibrational energy ($E_{\text{vib}}^{\text{mic}} \equiv E_{\text{kin}}^{\text{mic}} - Mv^2/2$), which for our purposes is determined by charge oscillations (e.g., dipole oscillations). Hence, it is possible to write the total kinetic and electromagnetic energies, $E_{\text{tot}}^{\text{mic}} \equiv \tilde{E}_{\text{tot}}^{\text{mic}} - Mc^2$, as:

$$E_{\text{tot}}^{\text{mic}} = E_{\text{EM}}^{\text{mic}} + E_{\text{vib}}^{\text{mic}} + \frac{1}{2}Mv^2. \tag{8}$$

In a few instances, we shall be interested in massive material bodies *at rest* with a mass that can be assumed infinitely large ($M \to \infty$) from the point of view of light interactions, so that the center of mass position can be regarded as time independent and the center of mass velocity is near zero, $\mathbf{v} \approx 0$, at all time instants. Importantly, in the described conditions the kinetic momentum $\mathbf{p}_{\text{kin}}^{\text{mic}} = M\mathbf{v}$ is generally finite and cannot be neglected (note that $\mathbf{v} \approx 0$ but $M \approx \infty$). In contrast, the center of mass kinetic energy can be safely disregarded ($\frac{1}{2}Mv^2 = \frac{1}{2}\mathbf{p}_{\text{kin}}^{\text{mic}} \cdot \mathbf{v} \approx 0$). For example, a mirror illuminated by an optical beam feels a continuous radiation pressure, and hence its kinetic momentum increases linearly with time. Nevertheless, if the mirror is sufficiently massive its position is nearly time independent and $\mathbf{v} \approx 0$.

For future reference, we note that for a closed system (in the absence of external forces or of dissipation) the following conservation law is necessarily satisfied at the microscopic level:

$$\nabla \cdot \mathbf{S}_{\text{tot}}^{\text{mic}} + \frac{d\tilde{W}_{\text{tot}}^{\text{mic}}}{dt} = 0. \tag{9}$$

Here, $\mathbf{S}_{\text{tot}}^{\text{mic}} = \mathbf{e} \times \mathbf{b} / \mu_0 + \mathbf{S}_{\text{mat}}^{\text{mic}}$ is the *total* microscopic energy density flux and $\tilde{W}_{\text{tot}}^{\text{mic}}$ is the total energy density at the microscopic level (including the rest mass energy) [2]. The energy density flux has a light component ($\mathbf{e} \times \mathbf{b} / \mu_0$) and a matter component ($\mathbf{S}_{\text{mat}}^{\text{mic}}$). The matter component $\mathbf{S}_{\text{mat}}^{\text{mic}}$ describes the energy flux determined by the flow of material particles, i.e., the flow of the rest mass energy and of the kinetic energy. It can be written as $\mathbf{S}_{\text{mat}}^{\text{mic}} = \rho_{\text{mat}} \left( c^2 + v_{\text{mic}}^2 / 2 \right) \mathbf{v}_{\text{mic}}$, where $\rho_{\text{mat}}$ is the mass density and $\mathbf{v}_{\text{mic}}$ is the microscopic velocity of the particles at point $\mathbf{r}$.

Within the $o\left(v_\alpha^2 / c^2\right)$ approximation, the relativistic relation between the energy and the momentum of a single particle ($\mathbf{p} = E\mathbf{v}/c^2$) is only satisfied with an accuracy $o\left(v_\alpha^2 / c^2\right)$. As a consequence, the relation between the matter momentum density ($\mathbf{g}_{\text{mat}}^{\text{mic}}$) and the matter energy density flux is $\mathbf{g}_{\text{mat}}^{\text{mic}} = \mathbf{S}_{\text{mat}}^{\text{mic}} / c^2$ only with precision $o\left(v_\alpha^2 / c^2\right)$. Indeed, within the previously discussed approximations one has $\mathbf{g}_{\text{mat}}^{\text{mic}} = \rho_{\text{mat}} \mathbf{v}_{\text{mic}}$ and therefore $\mathbf{g}_{\text{mat}}^{\text{mic}} = \mathbf{S}_{\text{mat}}^{\text{mic}} / c^2 + o\left(v_{\text{mic}}^2 / c^2\right)$. For this reason, the total momentum density (of light and matter) also satisfies $\mathbf{g}_{\text{tot}}^{\text{mic}} = \mathbf{S}_{\text{tot}}^{\text{mic}} / c^2 + o\left(v_{\text{mic}}^2 / c^2\right)$.

## C. *Limitations of macroscopic electrodynamics*

Macroscopic electrodynamics is a tremendously successful theory that reduces light-matter interaction problems of great complexity to simpler problems that are analytically (or numerically) tractable. This is done by averaging out the spatial fluctuations of the microscopic fields [15]. The increased simplicity necessarily implies that some information about the material is lost in the homogenization process. Nevertheless, effective medium theories are rather powerful and often enable us to predict with great accuracy the electrodynamics of complex systems. It is useful to revisit why this is so and highlight some limitations of macroscopic electrodynamics.

Let us suppose again that some material body is enclosed in a volume $V$ surrounded by air (or by a vacuum). It is assumed that the material response can be characterized in the framework of some effective medium theory that determines the time dynamics of the macroscopic fields $(\mathbf{E}, \mathbf{B})$. The only thing we need to assume for now is that the effective medium theory is good enough so that, for a given excitation, the macroscopic fields $(\mathbf{E}, \mathbf{B})$ agree in the air regions with the microscopic fields $(\mathbf{e}, \mathbf{b})$ determined by the same excitation. This property implies that with the macroscopic theory it is possible to determine both the Poynting vector and the stress-tensor in the air region, and that the corresponding formulas agree with the microscopic theory results. Hence, using simply the universal conservation laws for energy and momentum (and supposing that no material particles cross the boundary $\partial V$ in the time interval of interest) it possible to guarantee that:

$$\int_{\partial V} \mathbf{S}_0 \cdot \hat{\mathbf{n}} ds = -\frac{dE_{\text{tot}}^{\text{mic}}}{dt} - p_{\text{d}}^{\text{mic}}, \tag{10a}$$

$$\int_{\partial V} \overline{\mathbf{T}}_0 \cdot \hat{\mathbf{n}} ds = \frac{d}{dt}\left(\mathbf{p}_{\text{kin}}^{\text{mic}} + \mathbf{p}_{\text{EM}}^{\text{mic}}\right) - \mathbf{F}_{\text{ext}}^{\text{mic}}, \tag{10b}$$

with $\mathbf{S}_0 = \mathbf{E} \times \dfrac{\mathbf{B}}{\mu_0}$ and $\overline{\mathbf{T}}_0 = \varepsilon_0 \mathbf{E} \otimes \mathbf{E} + \mu_0^{-1} \mathbf{B} \otimes \mathbf{B} - \left(\dfrac{1}{2}\varepsilon_0 \mathbf{E} \cdot \mathbf{E} + \dfrac{1}{2}\mu_0^{-1} \mathbf{B} \cdot \mathbf{B}\right)\underline{\mathbf{1}}$. In the above, $E_{\text{tot}}^{\text{mic}}$ represents the total energy (of light and matter) enclosed by the volume $V$, and $\mathbf{p}_{\text{kin}}^{\text{mic}}$ is the kinetic momentum of the material particles in $V$. We allow the system to be open and interact with the exterior so that $p_{\text{d}}^{\text{mic}}$ represents the power dissipated (e.g., in the form of heat) and $\mathbf{F}_{\text{ext}}^{\text{mic}}$ represents the total force due to interactions with the exterior. All the quantities on the right-hand side of the Eqs. (10) are inherently microscopic, whereas the left-hand side is written in terms of the macroscopic electromagnetic fields.

Let us first consider that there is no dissipation ($p_{\text{d}}^{\text{mic}} = 0$, $\mathbf{F}_{\text{ext}}^{\text{mic}} = 0$). In this case, it is clear from (10) that it is possible within the framework of macroscopic electrodynamics to determine *exactly* how much is the total energy and total momentum stored in the material body. Note that both the energy and the momentum have light and matter components. For example, one may write (apart from a constant) $E_{\text{tot}}^{\text{mic}} = -\int_{-\infty}^{t}\int_{\partial V} \mathbf{S}_0 \cdot \hat{\mathbf{n}} dt ds$.

Usually one can do better than this. Often, in the framework of macroscopic electrodynamics it is possible for lossless systems to derive an energy conservation law of the form (this is possible even for dispersive systems, see Refs. [16-19]):

$$\nabla \cdot \mathbf{S}^{\text{mac}} + \frac{d}{dt}W^{\text{mac}} = 0, \tag{11}$$

where $\mathbf{S}^{\text{mac}}$ and $W^{\text{mac}}$ are some functions of the macroscopic fields (and possibly of other additional state variables that determine the macroscopic response of the medium), such that $\mathbf{S}^{\text{mac}} = \mathbf{S}_0$ in the air region. Note that the conservation law is supposed to hold everywhere in *V*, including at the boundaries between different regions. In that case, Eq. (10a) implies that $E_{tot}^{mic}$ can be written in terms of the macroscopic fields as (apart from an irrelevant constant):

$$E_{tot}^{mic} = \int_V W^{mac} dV. \tag{12}$$

Thus, $W^{mac}$ can be regarded as the macroscopic stored energy density and $\mathbf{S}^{mac}$ as the macroscopic Poynting vector. As an example, let us consider the case of a rigid material body characterized by some permittivity and permeability with negligible frequency variation. The body is assumed to be sufficiently massive ($M \to \infty$) so that its center of mass position is time independent and $\mathbf{v} \approx 0$. In that case, as is well known, $W^{mac} = \frac{1}{2}(\mathbf{D}\cdot\mathbf{E} + \mathbf{B}\cdot\mathbf{H})$ and $\mathbf{S}^{mac} = \mathbf{E}\times\mathbf{H}$ satisfy our requirements. Hence, $W^{mac}$ determines the macroscopic energy density associated with the electromagnetic energy and with the vibrational degrees of freedom of the material (i.e., with the dipole oscillations) such that in this example $E_{tot}^{mic} = E_{EM}^{mic} + E_{vib}^{mic}$ [20].

The situation changes completely in presence of material dissipation. Indeed, in such a case the right-hand side of Eq. (10a) has two terms, and even though their sum can be unequivocally determined with macroscopic electrodynamics, the individual parcels cannot. Indeed, in dissipative media it is *impossible* to determine unambiguously the electromagnetic energy stored in a medium using a purely macroscopic theory, i.e., with no detailed knowledge of the microscopic mechanisms that lead to the dissipation [21]. Note that if one can find some decomposition of the type $\nabla \cdot \mathbf{S}^{mac} + \frac{d}{dt}W^{mac} + q^{mac} = 0$ ($\mathbf{S}^{mac}$, $W^{mac}$, $q^{mac}$ being some functions of the macroscopic fields, and $\mathbf{S}^{mac} = \mathbf{S}_0$ in the air region) then one may write $\frac{d}{dt}E_{\text{tot}}^{\text{mic}} + p_{\text{d}}^{\text{mic}} = \frac{d}{dt}\int_V W^{\text{mac}}dV + \int_V q^{\text{mac}}dV$. However, in general it is abusive and incorrect to identify $p_{\text{d}}^{\text{mic}} = \int_V q^{\text{mac}}dV$ and $E_{\text{tot}}^{\text{mic}} = \int_V W^{\text{mac}}dV$. Indeed, the first objection is that $\mathbf{S}^{\text{mac}}$, $W^{\text{mac}}$, $q^{\text{mac}}$ may not be uniquely defined and that one may find alternative $\mathbf{S}'^{\text{mac}}$, $W'^{\text{mac}}$, $q'^{\text{mac}}$ that lead to *different* expressions for the stored energy and dissipated power. A more serious argument that demonstrates the impossibility of writing $E_{\text{tot}}^{\text{mic}}$ in terms of the macroscopic fields is illustrated by the following example.

Consider some artificial material (metamaterial) formed by an array of loaded dipoles. It is supposed that each inclusion is formed by two metallic cylinders joined by a lumped load (similar to a short dipole antenna with a load at the terminals). For long wavelengths, the metamaterial response can be described using effective medium methods (see e.g., [22]). We consider two possibilities for the load: *(i)* a simple resistor $R$, *(ii)* the Vainshtein circuit [21]. The Vainshtein circuit consists of two branches connected in parallel. The first branch is the series of a resistor $R$ with an inductor $L$, and the second branch is the series of another resistor $R$ with a capacitor $C$, with $L/C = R^2$. Interestingly, the input impedance of the Vainshtein circuit, defined consistently with $\frac{1}{Z_{in}} = \frac{1}{R-i\omega L} + \frac{1}{R+1/(-i\omega C)}$, is simply $R$ for all frequencies. Clearly, the metamaterial interacts with an external excitation *exactly* in the same way independent if the load is *(i)* or *(ii)*, and thus the effective medium response is independent whether the load is a simple resistor $R$ or the corresponding Vainshtein circuit. However, the two loads are fundamentally different: one is purely dissipative and the other is able to store electromagnetic energy within it. Clearly, an effective medium theory is not able to distinguish between the two cases, and hence it is fundamentally impossible to determine the instantaneous stored energy or the instantaneous dissipation rate with macroscopic electrodynamics in presence of material losses. Thus, macroscopic theories have intrinsic limitations, and that is the inevitable penalty that comes with a simplified description of wave phenomena.

These problems are partially alleviated in time-harmonic regime. In such a case, integrating $\frac{d}{dt}E_{\text{tot}}^{\text{mic}} + p_{\text{d}}^{\text{mic}} = \frac{d}{dt}\int_V W^{\text{mac}}dV + \int_V q^{\text{mac}}dV$ over one time period gives: $\langle p_{\text{d}}^{\text{mic}}\rangle_T = \int_V \langle q^{\text{mac}}\rangle_T dV$. Hence, the time-averaged dissipation rate can be determined unequivocally using macroscopic electrodynamics [21, 23, 24].

## III. The Electromagnetic Momentum

### A. Link between the light momentum and the macroscopic fields

Unfortunately, Eq. (10b) does not help in any way to decide which is the correct form of the kinetic electromagnetic momentum, even for a closed system ($\mathbf{F}_{\text{ext}}^{\text{mic}} = 0$). In short, the problem is that within a macroscopic formalism one knows the total instantaneous momentum that is transferred to the material ($\mathbf{p}_{\text{kin}}^{\text{mic}} + \mathbf{p}_{\text{EM}}^{\text{mic}}$) but it is impossible to tell which part of the transferred momentum goes to the optical field in the material and which part goes to the kinetic momentum of the material particles. Note that this limitation applies as well to massive bodies ($M \to \infty$) at rest, because as discussed previously the transferred kinetic momentum cannot be ignored even if the center of mass position is assumed time-independent. In contrast, as demonstrated in Sect. IIC, in the absence of dissipation Eq. (10a) allows one to precisely determine the instantaneous value of the energy $E_{EM}^{mic} + E_{vib}^{mic}$ because for massive bodies the center of mass kinetic energy ($\frac{1}{2}Mv^2$) is negligible. Clearly, macroscopic electrodynamics has more severe limitations in the context of the momentum problem than in the stored energy problem.

Remarkably, the energy conservation law (9) can shed some light on the resolution of the puzzle. Let us suppose again that the material is lossless so that the microscopic conservation law (9) holds. Furthermore, let $\mathbf{S}^{mac}$ be some function of the macroscopic fields with $\mathbf{S}^{mac} = \mathbf{S}_0$ in the air regions and such that $\mathbf{S}^{mac}$ satisfies Eq. (11) for some function of the macroscopic fields $W^{mac}$. Supposing again that in the air region the macroscopic fields can be identified with the microscopic fields and that no material particles cross the volume boundary, it is possible to write:

$$\frac{1}{c^2}\int_{\partial V}\mathbf{r}\left(\mathbf{S}^{mac}\cdot\hat{\mathbf{n}}\right)ds = \frac{1}{c^2}\int_{\partial V}\mathbf{r}\left(\mathbf{e}\times\frac{\mathbf{b}}{\mu_0}\right)\cdot\hat{\mathbf{n}}ds = \frac{1}{c^2}\int_{\partial V}\mathbf{r}\left(\mathbf{S}_{tot}^{mic}\cdot\hat{\mathbf{n}}\right)ds, \qquad (13)$$

where $\mathbf{r} = (x, y, z)$ is the position vector. But from the divergence theorem and Eqs. (9) and (11) one finds that:

$$\frac{1}{c^2}\int_V \mathbf{S}_{mic}^{tot}dV = \frac{1}{c^2}\int_V \mathbf{S}^{mac}dV + \frac{d\tilde{\Theta}}{dt}, \qquad (14)$$

where we introduced

$$\tilde{\Theta} = \frac{1}{c^2}\int_V \mathbf{r}\left(\tilde{W}_{tot}^{mic} - W^{mac}\right)dV. \qquad (15)$$

As discussed in the end of Sect. II.B, $\mathbf{S}_{tot}^{mic}/c^2$ coincides with the total microscopic momentum density. In particular, we can write $\mathbf{p}_{tot}^{mic} = \frac{1}{c^2}\int_V \mathbf{S}_{tot}^{mic}ds$ with $\mathbf{p}_{tot}^{mic} = \mathbf{p}_{EM}^{mic} + M\mathbf{v}$ [see Eq. (6)] with an accuracy $o(v_\alpha^2/c^2)$. It is also useful to decompose the total energy density as $\tilde{W}_{tot}^{mic} = W_{tot}^{mic} + c^2 \rho_{mat}$, such that the second term is the contribution of the rest mass energy. Taking into account that $\frac{d}{dt}\int_V \mathbf{r}\rho_{mat} dV = M\mathbf{v}$, we obtain the key result:

$$\mathbf{p}_{EM}^{mic} = \frac{1}{c^2}\int_V \mathbf{S}^{mac} dV + \frac{d\mathbf{\Theta}}{dt}. \tag{16}$$

where,

$$\mathbf{\Theta} = \frac{1}{c^2}\int_V \mathbf{r}\left(W_{tot}^{mic} - W^{mac}\right) dV. \tag{17}$$

Note that $W_{tot}^{mic}$ stands for the total energy density excluding the rest mass energy. We would like to highlight that the unique hypotheses used in the derivation of Eq. (16) are *(i)* the microscopic and macroscopic fields agree outside the material system and *(ii)* the conservation of energy (material is lossless).

## B. Steady-state regime

The first term in the right-hand side of Eq. (16) depends exclusively on the macroscopic fields. Unfortunately, the second term depends on both the macroscopic fields and on the unknown microscopic fields. Importantly, in many physical situations one can assume that the system is in a stationary state. The simplest example is when the electromagnetic excitation is periodic in time (e.g., time harmonic excitation) and all the material bodies are at rest, e.g., they are very massive so that their center of mass position is time independent and $\mathbf{v} \approx 0$. As discussed in Sect. IIB, in the $M \to \infty$ limit the kinetic momentum $\mathbf{p}_{kin}^{mic} = M\mathbf{v}$ is finite and may vary with time to account for the momentum transferred by the Lorentz force, but the center of mass kinetic energy is negligibly small $\frac{1}{2}\mathbf{p}_{\text{kin}}^{\text{mic}} \cdot \mathbf{v} \approx 0$. In the outlined scenario, both $W_{\text{tot}}^{\text{mic}}$ and $W^{\text{mac}}$ are periodic in time and hence the time average of the second term in Eq. (18) vanishes. Thus, it follows that:

$$\left\langle \mathbf{p}_{\text{EM}}^{\text{mic}} \right\rangle_T = \frac{1}{c^2} \int_V \left\langle \mathbf{S}^{\text{mac}} \right\rangle_T dV. \tag{18}$$

Therefore, the time-averaged electromagnetic momentum can be exactly calculated using macroscopic electrodynamics when $\mathbf{v} \approx 0$. In the particular case of *local* media (e.g., for any lossless dispersive bianisotropic material modeled by local in space constitutive relations; see Ref. [16]) it is well known that Eq. (11) is satisfied with $\mathbf{S}^{\text{mac}} = \mathbf{E} \times \mathbf{H}$. Thus, for local media one has:

$$\left\langle \mathbf{p}_{\text{EM}}^{\text{mic}} \right\rangle_T = \int_V \left\langle \mathbf{g}_{\text{EM}}^{\text{Ab}} \right\rangle_T dV, \qquad \text{(local materials)}. \tag{19}$$

Thus, in the enunciated conditions the time-averaged (kinetic) electromagnetic momentum agrees exactly with the Abraham form. These arguments enable us to reject the Minkowski momentum as a valid kinetic electromagnetic momentum for light (in the sense that it cannot predict $\int_V \varepsilon_0 \mathbf{e} \times \mathbf{b} dV$). This conclusion is consistent with the modern understanding that the Minkowski momentum is instead the canonical light momentum [13]. However, Eq. (18) also demonstrates that the Abraham form cannot be universal. Indeed, it is well known that for nonlocal materials in general the macroscopic Poynting vector $\mathbf{S}^{\text{mac}}$ *cannot* be identified with $\mathbf{E} \times \mathbf{H}$ [21] (see Ref. [25] for a specific example). Thus, the Abraham form of the electromagnetic momentum is only acceptable in local media (i.e., media that have a local response in the co-moving frame). Crucially, the results (18) and (19) agree exactly with what was found for a bulk medium in Refs. [23,

## C. Transient regime

Even though Eq. (18) is an important step forward in the understanding of the problem, it is not the final word. Indeed, the instantaneous Lorentz force depends on the detailed variation in time of $\mathbf{p}_{EM}^{mic}$ [see Eq. (2)], which is given by Eq. (16), and relies on the unknown microscopic fields through the $\Theta$ vector function. Here, we note that the integral that determines $\Theta$ is independent of the origin of the coordinate system because from Eq. (12) one has $\int_V \left( W_{tot}^{mic} - W^{mac} \right) dV = 0$.

Unfortunately, without further knowledge of the microstructure, it does not seem possible to write $\Theta$ using only the macroscopic fields. This appears to be another intrinsic limitation of macroscopic electrodynamics. So it is not surprising that the Abraham-Minkowski controversy has lasted for over a century: it seems that without further assumptions the time dynamics of $\mathbf{p}_{EM}^{mic}$ cannot be *exactly* predicted within the realm of macroscopic electrodynamics.

Nevertheless, one can decompose the volume $V$ into its elementary basic cells (it is supposed without loss of generality that the material has a crystalline structure; the *m*-th cell is denoted by $\Omega_m$) to obtain:

$$\begin{aligned}\Theta &= \frac{1}{c^2} \sum_m \int_{\Omega_m} \mathbf{r} \left( W_{tot}^{mic} - W^{mac} \right) dV \\ &= \frac{1}{c^2} \sum_m \mathbf{r}_m \int_{\Omega_m} \left( W_{tot}^{mic} - W^{mac} \right) dV + \frac{1}{c^2} \sum_m \int_{\Omega_m} \delta \mathbf{r}_m \left( W_{tot}^{mic} - W^{mac} \right) dV \end{aligned} \quad (20)$$

$\mathbf{r}_m$ being the center of the *m*-th cell and $\delta\mathbf{r}_m = \mathbf{r} - \mathbf{r}_m$. Any reasonable homogenization theory should ensure that when the material body is at rest $\int_{\Omega_m}\left(W_{tot}^{mic} - W^{mac}\right)dV \approx 0$, i.e. it should guarantee that *inside* the material $W^{mac}$ corresponds to the spatially averaged microscopic energy density [23, 24]. Thus, it is possible to estimate that $\mathbf{\Theta} \approx \frac{1}{c^2}\sum_m \int_{\Omega_m} \delta\mathbf{r}_m \left(W_{tot}^{mic} - W^{mac}\right)dV$ so that from Schwarz's inequality a generic (*l*-th) component of $\mathbf{\Theta}$ satisfies:

$$|\Theta_l| \leq \frac{a}{c^2}\sqrt{\frac{V}{12}\int_V \left(W_{tot}^{mic} - W^{mac}\right)^2 dV} \qquad (21)$$

where $V = \int_V dV$ is the volume of the relevant body. To obtain the upper bound it was assumed for simplicity that the lattice is cubic and has a period *a*. From this result, we can very roughly estimate (taking $W_{tot}^{mic} - W^{mac} \sim S^{mac}/c$ and considering $S^{mac}$ uniform within the material) that the relative difference of the magnitudes of second and first terms of Eq. (16) is on the order of $\omega_{max}a/c$. Here, $\omega_{max}$ is the maximum frequency of interest, i.e., it determines the largest frequency of the electromagnetic signal spectrum. Therefore, these arguments indicate that for material bodies at rest:

$$\mathbf{p}_{EM}^{mic} = \left(\frac{1}{c^2}\int_V \mathbf{S}^{mac} dV\right)\left[1 + o\left(\frac{\omega_{max}a}{c}\right)\right]. \qquad (22)$$

Thus, it seems that albeit nonzero, the vector $d\mathbf{\Theta}/dt$ may be negligible as compared to the first term in Eq. (16). The approximation $d\mathbf{\Theta}/dt \approx 0$ is expected to be particularly good for local media and when the lattice constant is deeply subwavelength as in natural materials.

In rigor, our theory only applies to non-dissipative materials. Evidently, if the Abraham expression of the light momentum is assumed generally valid (e.g., for local media this seems to be a reasonable approximation) the electromagnetic theory becomes complete and can be used to fully characterize arbitrary light-matter interactions even in presence of dissipation (see for example Ref. [26]).

## *D. Quantum optics*

It is interesting to discuss the consequences of our findings in the context of quantum optics. To this end, let us consider that the state of the relevant material system is stationary, i.e. it is an eigenstate of the Hamiltonian. The results of Sect. IIIB demonstrate unequivocally that in the framework of macroscopic electrodynamics the light momentum of a stationary state is determined by the operator (the "hat" indicates that the relevant symbol should be regarded as an operator):

$$\hat{\mathbf{p}}_{EM}^{mic} = \frac{1}{2c^2} \int_V \hat{\mathbf{E}} \times \hat{\mathbf{H}} - \hat{\mathbf{H}} \times \hat{\mathbf{E}} \, dV, \quad \text{(stationary state)}. \tag{23}$$

The above result applies to dispersive (possibly inhomogeneous) local media at rest with negligible material absorption. We used the fact that the expectation of $\frac{d\hat{\mathbf{\Theta}}}{dt}$ vanishes when the system is in an energy stationary state. The quantization of dispersive and nondispersive material platforms is discussed in Refs. [27-30].

One of the arguments typically enunciated in the literature in favor of the Minkowski momentum is that for a photon, i.e., for a light quantum, it predicts that the light momentum is $\hbar\mathbf{k}$ [2, 5]. To analyze in detail this idea let us consider some closed system invariant to translations along the *x*-direction. For definiteness, it may be a cavity formed by nondispersive material slabs (stacked along the *z*-direction) and terminated with periodic boundaries in the walls perpendicular to the *x*-direction. When the system is invariant to translations along *x* it is well-known that $k_x$ is a good quantum number. For nondispersive materials the Minkowski momentum operator, $\hat{\mathbf{p}}_w = \frac{1}{2}\int \hat{\mathbf{D}} \times \hat{\mathbf{B}} - \hat{\mathbf{B}} \times \hat{\mathbf{D}}\, dV$ (here the integration is over the entire cavity), has an expectation such that

$$\left\langle m_{k_x} | \hat{\mathbf{x}} \cdot \hat{\mathbf{p}}_w | m_{k_x} \right\rangle = \hbar k_x \left( m + \frac{1}{2} \right). \tag{24}$$

where $|m_{k_x}\rangle$ is an energy state with *m* quanta in the field mode with label $k_x$ [20, Ap. B]. Hence, when a single quantum is created the Minkowski momentum along the *x* direction increases by the amount $\hbar k_x$. Evidently, because $\hat{\mathbf{p}}_w \neq \hat{\mathbf{p}}_{EM}^{mic}$ the momentum "$\hbar k_x$" cannot be understood as the "kinetic" momentum of a photon. The reason is simple to understand: due to the light-matter interactions the field quanta are "dressed" particles (quasiparticles) corresponding to the hybridization of the elementary light and matter degrees of freedom [14].

To further elaborate on this idea, let us consider the general formula (2) for the Lorentz force. When the system is invariant to translations along the *x*-direction it is possible to show that $\int_{\partial V} \hat{\mathbf{n}} \cdot \overline{\mathbf{T}} \cdot \hat{\mathbf{x}}\, ds = \hat{\mathbf{x}} \cdot \frac{d\mathbf{p}_w}{dt}$ [31, Ap. A]. Therefore, the (*x*-component) of the Lorentz force operator acting on the *i*-th material slab in the cavity is:

$$\hat{F}_{L,i}^{mic} = -\hat{\mathbf{x}} \cdot \frac{d}{dt}\hat{\mathbf{p}}_{EM,i}^{mic} + \hat{\mathbf{x}} \cdot \frac{d}{dt}\hat{\mathbf{p}}_{w,i}. \tag{25}$$

The right-side of the equation is often known as the Abraham force [2, 11]. It is implicit that the momentum operators are associated with integrals over the volume $V_i$ of the *i*-th material slab. Here, we adopt the Schrödinger picture so that it is implicit that $\frac{d\hat{\mathbf{p}}_{w,i}}{dt}$ stands for the operator $\frac{1}{i\hbar}[\hat{\mathbf{p}}_{w,i}, \hat{H}]$, $\hat{H}$ being the Hamiltonian of the system. Evidently, for an energy eigenstate the expectation of the Lorentz force along $x$ vanishes. Let us imagine a situation wherein the system is initially ($t<0$) in a stationary state, let us say in the quantum vacuum state $|i\rangle=|0\rangle$, then during some period of time is externally excited (e.g., with some current source so that $\hat{H}=\hat{H}(t)$), and then the excitation is switched off so that for $t>t_0$ the system is left in the final $|f\rangle=|1_{k_x}\rangle$ state. The total kinetic momentum transferred to the $i$-th material slab by the excitation is

$$\delta p_{\text{kin},i} \equiv \int_0^{t_0} \hat{F}_{\text{L},i}^{\text{mic}} dt = \langle f | \hat{\mathbf{x}} \cdot (\hat{\mathbf{p}}_{w,i} - \hat{\mathbf{p}}_{\text{EM},i}^{\text{mic}}) | f \rangle - \langle i | \hat{\mathbf{x}} \cdot (\hat{\mathbf{p}}_{w,i} - \hat{\mathbf{p}}_{\text{EM},i}^{\text{mic}}) | i \rangle$$
$$= \langle f | \hat{\mathbf{x}} \cdot \hat{\mathbf{p}}_{\text{ps},i} | f \rangle - \langle i | \hat{\mathbf{x}} \cdot \hat{\mathbf{p}}_{\text{ps},i} | i \rangle$$
(26)

In the above, $\hat{\mathbf{p}}_{\text{ps},i} = \frac{1}{2}\int_{V_i}\left(\hat{\mathbf{D}}\times\hat{\mathbf{B}} - \frac{1}{c^2}\hat{\mathbf{E}}\times\hat{\mathbf{H}}\right)dV + H.c.$ is the so-called pseudo-momentum operator associated with the $i$-th material slab [32], and in the second identity we used the fact that the electromagnetic momentum is determined by the Abraham momentum for the initial and final states [Eq. (23)]. Hence, the excitation of the quantum system leads to a change in the kinetic momentum of the $i$-th slab ($\delta p_{\text{kin},i}$) that is uniquely determined by the expectation of the pseudo-momentum in the final and initial states, and quite remarkably is totally independent of the details of the excitation. It is now simple to verify that Eqs. (24) and (26) imply that $\hbar k_x = \sum_i \delta p_{\text{kin},i} + \langle f | \hat{\mathbf{x}} \cdot \hat{\mathbf{p}}_{\text{EM}}^{\text{mic}} | f \rangle - \langle i | \hat{\mathbf{x}} \cdot \hat{\mathbf{p}}_{\text{EM}}^{\text{mic}} | i \rangle$ where $\hat{\mathbf{x}} \cdot \hat{\mathbf{p}}_{\text{EM}}^{\text{mic}}$ stands for the operator associated with the total electromagnetic momentum of the cavity. This confirms that the Minkowski (canonical) momentum ($\hbar k_x$) has both light and matter components, and that the state $|1_{k_x}\rangle$ is indeed associated with a quasiparticle. It is underscored that our result relies exclusively on the simple assumptions enunciated in the end of Sect. IIIA.

## IV. Moving Media

The case of moving material bodies with an arbitrary shape is considerably more difficult because usually it is not simple to find a conservation law of the form (11) in the framework of macroscopic electromagnetism (i.e., with $W^{\text{mac}}, \mathbf{S}^{\text{mac}}$ functions only of the macroscopic fields). The reason is that the light-matter interactions usually originate the conversion of kinetic energy into light, and vice-versa. In particular, a moving material can gain (or lose) kinetic energy due to the optical forces, and from an electromagnetic point of view this corresponds to a form of dissipation (or gain).

Nevertheless, for small velocities the corrections on the electromagnetic momentum must be on the order of $\mathbf{v}/c$, $\mathbf{v}$ being the center of mass velocity of the relevant body. For example, for a moving material with a deeply subwavelength lattice period and a local response in the co-moving frame, it follows from Eq. (22) that:

$$\mathbf{p}_{\text{EM}}^{\text{mic}} = \frac{1}{c^2} \int_V \mathbf{E} \times \mathbf{H} dV + o(v/c). \tag{27}$$

From Eq. (2) the Lorentz force acting on the moving body can also be determined with a similar accuracy:

$$\mathbf{F}_{\text{L}}^{\text{mic}} = -\frac{1}{c^2} \frac{d}{dt} \int_V \mathbf{E} \times \mathbf{H} dV + \int_{\partial V} \hat{\mathbf{n}} \cdot \overline{\mathbf{T}} ds + o(v/c). \tag{28}$$

In the following, we shall prove that for nondispersive moving media in steady-state conditions the previous results can be made more precise.

## A. Nondispersive moving media

Next, we consider conventional dielectric media with negligible material dispersion, such that the permittivity $\varepsilon$ and the permeability $\mu$ are independent of frequency in the co-moving frame wherein the material is at rest. It is well known, that in a generic inertial frame (e.g., the laboratory frame) the material response is bianisotropic [33, 34, 35, 36] and is described by the following material matrix within a $o(v^2/c^2)$ approximation [34, Sec. 76]:

$$\mathbf{M} = \begin{pmatrix} \varepsilon_0 \varepsilon \mathbf{1}_{3\times 3} & \dfrac{(\varepsilon\mu - 1)}{c^2} \mathbf{v} \times \mathbf{1}_{3\times 3} \\ -\dfrac{(\varepsilon\mu - 1)}{c^2} \mathbf{v} \times \mathbf{1}_{3\times 3} & \mu_0 \mu \mathbf{1}_{3\times 3} \end{pmatrix}. \tag{29}$$

The material matrix relates the electromagnetic fields in the laboratory frame as $\begin{pmatrix} \mathbf{D} \\ \mathbf{B} \end{pmatrix} = \mathbf{M} \cdot \begin{pmatrix} \mathbf{E} \\ \mathbf{H} \end{pmatrix}$. Here, $\mathbf{v}$ is the velocity of the moving dielectric with respect to the laboratory frame.

We will focus in the case wherein the material boundaries do not change with time and the flow of matter is stationary so that $\mathbf{v} = \mathbf{v}(\mathbf{r})$ at a given point of space can be assumed time independent. Thus, the matter flow can be depicted as a flow in closed orbits, i.e., it corresponds to a "circulatory flow". The simplest example is the case of a rigid cylindrical body under a rotational motion about the symmetry axis. However the notion of "circulatory flow" is more general and does not have to be associated with a rotation of a rigid body, for example it applies as well to fluids. Note that if the flow of matter is stationary in laboratory frame it follows that the center of mass velocity vanishes.

## B. Electromagnetic momentum

The analysis of Sect. IIIA remains rigorously valid when applied to a generic nondispersive circulatory flow. In particular Eq. (16) still holds in the present context. For nondispersive media [such that $\nabla \times \mathbf{E} = -\partial_t \mathbf{B}$ and $\nabla \times \mathbf{H} = \partial_t \mathbf{D}$, with the electromagnetic fields linked by a symmetric material matrix $\mathbf{M} = \mathbf{M}(\mathbf{r})$ defined as in Eq. (29)], it is easy to check that the macroscopic fields satisfy

$$\nabla \cdot (\mathbf{E} \times \mathbf{H}) + \frac{d}{dt} W_w = 0, \tag{30}$$

where $W_w = \frac{1}{2}(\mathbf{D} \cdot \mathbf{E} + \mathbf{B} \cdot \mathbf{H})$ is the wave energy density whose physical meaning in moving media is further discussed in Appendix A. The above equation has the same structure as Eq. (11) and holds in all space (including at the boundaries). Hence, it follows from Eq. (16) that for a circulatory flow the electromagnetic momentum is given by $\mathbf{p}_{EM}^{mic} = \frac{1}{c^2} \int_V \mathbf{E} \times \mathbf{H} \, dV + \frac{d\mathbf{\Theta}}{dt}$ with $\mathbf{\Theta} = \frac{1}{c^2} \int_V \mathbf{r} \left( W_{tot}^{mic} - W_w \right) dV$. Then, in a stationary regime (e.g., for time harmonic excitation) the time averaged electromagnetic momentum is rigorously determined by the Abraham momentum density (up to corrections on the order of $o(v_\alpha^2/c^2)$):

$$\left\langle \mathbf{p}_{EM}^{mic} \right\rangle_T = \frac{1}{c^2} \int_V \left\langle \mathbf{E} \times \mathbf{H} \right\rangle_T dV, \qquad \text{(circulatory flow).} \tag{31}$$

Therefore, quite remarkably, for nondispersive circulatory flows the Abraham momentum also determines *exactly* the light momentum in steady-state conditions. Note that it is implicit that in the stationary regime the time variation of the velocity $\mathbf{v}(\mathbf{r})$ is negligible, and in particular the center of mass velocity remains approximately zero. For the same reasons as in Sect. IIIA, the changes in the kinetic momentum cannot be neglected.

Unfortunately, it is not obvious how to generalize Eq. (30) and thereby Eq. (31) to the case of dispersive media (at least in a general case). The problem is that the response of a dispersive dielectric depends on its past history, and when the material is moving this implies that the response depends on the trajectory travelled by the material elements. In other words, the response is spatially dispersive. Besides that, lossless dispersive media in motion are generically "unstable" in the sense that a stationary state may be impossible to attain. Indeed, a dispersive moving system generically supports electromagnetic instabilities that lead to the spontaneous conversion of kinetic energy into electromagnetic energy. For more details the reader is referred to Refs. [30, 31, 37-40].

## C. Quantum optics in moving systems

To illustrate the consequences of Eq. (31) in the context of quantum optics, first we note that it implies that the electromagnetic momentum of an energy eigenstate can be determined with the operator (23), exactly as when all the materials are at rest. The quantization of the electromagnetic field in moving material platforms is discussed in Refs. [20, 31, 35].

The result (31) holds rigorously for a circulatory flow of matter, but may be extended to the case of material slabs with a constant velocity. To show this, we start with the configuration of Fig. 1a, which corresponds to a generic circulatory flow and for which Eq. (23) holds. If the end-effects are neglected (corresponding to the regions where the velocity $\mathbf{v} = \mathbf{v}(\mathbf{r})$ changes continuously from $\mathbf{v} = v\hat{\mathbf{x}}$ to $\mathbf{v} = -v\hat{\mathbf{x}}$) it seems reasonable to replace the original configuration by that of Fig. 1b where the lateral walls (perpendicular to the *x*-direction) are taken as periodic boundaries, so that the system becomes invariant to translations along the *x*-direction. Thus, the geometry of Fig. 1b may be regarded as a simpler mathematical model of the circulatory flow, and with that understanding it is acceptable to use Eq. (23) in the scenario of Fig. 1b. Note that the two dielectrics with velocities $\mathbf{v}=v\hat{\mathbf{x}}$ and $\mathbf{v}=-v\hat{\mathbf{x}}$ may be separated by an arbitrary material system at rest, for example by a metal plate, and in that case the top and bottom regions are effectively screened and do not interact.

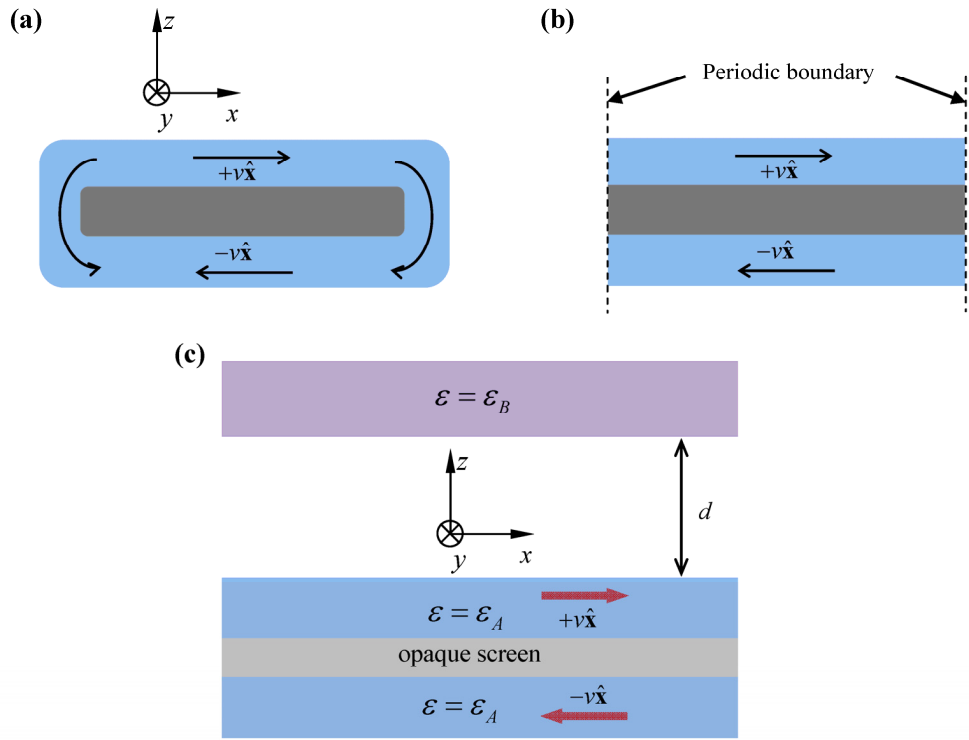

Fig. 1. (a) A generic circulatory flow of polarizable neutral matter. (b) Simplified mathematical model for the circulatory flow, where the end-effects are modeled with periodic boundaries. (c) A cavity formed by two interacting systems *A* and *B* separated by a distance *d*. The system B is at rest, while the system *A* corresponds to a circulatory flow of polarizable neutral matter.

Let us now concentrate on the scenario of Fig. 1c where a nondispersive dielectric slab at rest (system B) is separated by a nondispersive circulatory flow (system A) by a

-28-

vacuum gap with thickness *d*. It is implicit that the cavity walls normal to the *x*-direction are terminated with periodic boundaries, similar to Fig. 1b.

Importantly, in the described scenario the (*x*-component) of the Lorentz force operator is still determined by Eq. (25) (see Ref. [20] for the expression of the quantized field operators), because for moving media it remains true that $\int_{\partial V} \hat{\mathbf{n}} \cdot \overline{\mathbf{T}} \cdot \hat{\mathbf{x}} \, ds = \hat{\mathbf{x}} \cdot \frac{d\mathbf{p}_w}{dt}$ [31, Ap. A]. In particular, for an energy stationary state the expectation of the Lorentz force vanishes and there is no quantum friction, i.e., the expectation of the lateral optical force vanishes [41]. This conclusion is valid as long as the system state is truly stationary; as discussed in Appendix A, this is possible only if the relative velocity between the material bodies does not exceed a threshold limit.

Remarkably, if the quantum system is perturbed somehow the expectation of the lateral force can be nonzero, and the kinetic momentum transferred to the materials during the perturbation can be calculated using the pseudo-momentum operator [Eq. (26)]. To illustrate this, let us consider the interesting situation in which the distance between the dielectric slab at rest and the material body associated with the circulatory flow is initially very large $d \to \infty$. Supposing that the quantum system is in the ground state it is clear that by symmetry $\langle i | \hat{\mathbf{x}} \cdot \hat{\mathbf{p}}_{ps,l} | i \rangle = 0$ in this initial state (*l*=A,B). Let us suppose that the perturbation is such that the subsystems *A* and *B* are brought together (e.g., by a mechanical force directed along *z*) so that in the final state the distance *d* is finite. The process is assumed adiabatic so that the quantum system is brought from the initial to the final state through the parametric ground state $|0_d\rangle$ (the ground state depends on the distance *d* between the bodies). Then, according to Eq. (26) in this process there is a kinetic momentum transfer determined by $\delta p_{\text{kin},l} = \langle f | \hat{\mathbf{x}} \cdot \hat{\mathbf{p}}_{\text{ps},l} | f \rangle$ for each sub-system ($l$=A,B). The described phenomenon is some kind of quantum Hall effect, because the external force is applied along *z* but it induces a lateral optical force along *x*. This theory was developed in detail in Ref. [20] where it was demonstrated that $\delta p_{\text{kin},l}$ can be numerically determined using the zero-point Casimir interaction energy. Furthermore, it was shown that $\delta p_{\text{kin},A} + \delta p_{\text{kin},B} = 0$, i.e., the total change in the kinetic momentum is zero, so that the effect corresponds to an exchange of kinetic momentum by two material systems induced by the quantum fluctuations of the vacuum. Note that for sufficiently massive bodies the kinetic energy transfer has negligible consequences in the center of mass position of each subsystem. It is relevant to mention that the described effect may lead to a circulation of the quantum vacuum momentum in closed orbits, an effect that recently raised some attention in the literature [42-43].

## V. Conclusion

In summary, we revisited the Abraham-Minkowski controversy on the definition of the light momentum in a macroscopic material. Our point of view of this problem, which we feel is not sufficiently highlighted in the most influential articles of the recent literature, is that the dilemma has no universal resolution due to the intrinsic limitations of macroscopic electrodynamics. It was emphasized that the use of effective medium theories requires an inevitable compromise between simplicity and rigor, and invariably implies some loss of information about the state of the system, e.g., on the stored energy and on the stored momentum. Despite these fundamental limitations, it was shown that provided *(i)* the material absorption is negligible and *(ii)* the macroscopic theory accurately predicts the dynamics of the electromagnetic fields in the air regions, then the kinetic light momentum can be determined unequivocally under steady-state conditions for either material bodies at rest or for circulatory flows of neutral matter. In particular, it was shown that for local media the kinetic light momentum is determined by the Abraham result under steady-state conditions. In transient regime it seems impossible to determine exactly the stored light momentum in general conditions, but it was shown that for local media with ultra-subwavelength nanoscopic constituents the Abraham result is expected to approximate well the instantaneous kinetic light momentum. Finally, we discussed the implications of our findings in the context of quantum optics, highlighting that for energy stationary states the expectation of the kinetic electromagnetic momentum can be exactly determined with the Abraham formula. Furthermore, it was illustrated with a simple example that the canonical momentum of a light quantum in a medium ($\hbar \mathbf{k}$) has a matter component.

## Appendix A: The wave energy in moving media

In this Appendix, we discuss the physical meaning of the wave energy $W_\mathrm{w} = \frac{1}{2}(\mathbf{D}\cdot\mathbf{E} + \mathbf{B}\cdot\mathbf{H})$ in moving media. As mentioned in Sect. IIC, for media at rest it can be regarded as the stored energy density associated with the electromagnetic field and with the vibrational kinetic energies ($E_\mathrm{EM}^\mathrm{mic} + E_\mathrm{vib}^\mathrm{mic}$). Next, we refer to the energy density determined by $E_\mathrm{EM}^\mathrm{mic} + E_\mathrm{vib}^\mathrm{mic}$ as $W_\mathrm{EM+vib}$. Importantly, it was shown in Ref. [37, Ap. A] that for moving media $W_\mathrm{EM+vib} \neq W_\mathrm{w}$. The correct relation is $W_\mathrm{EM+vib} = W_\mathrm{w} - \mathbf{v}\cdot\mathbf{g}_\mathrm{ps}$ where

$$\mathbf{g}_\mathrm{ps} = \mathbf{D}\times\mathbf{B} - \frac{1}{c^2}\mathbf{E}\times\mathbf{H}, \tag{A1}$$

is the pseudo-momentum density. To better illustrate the concept, let us consider the geometry of Fig. 1c. Integrating the wave energy density over the top moving slab of subsystem $A$ (with velocity $\mathbf{v} = +v\hat{\mathbf{x}}$) it is found that $E_{\text{EM},A}^{\text{mic}} + E_{\text{vib},A}^{\text{mic}} = E_{\text{w},A} - \mathbf{v}\cdot\mathbf{p}_{\text{ps},A}$. Thus, for moving nondispersive materials the stored energy is determined by *both* the wave energy and the pseudo-momentum [31, 37]. In particular, the wave energy is not required to be positive, i.e., the wave energy can be negative, even though $E_{\text{EM}}^{\text{mic}} + E_{\text{vib}}^{\text{mic}}$ is always positive for moving dielectrics [31, 37]. Indeed, in can be checked that the material matrix $\mathbf{M}$ in Eq. (29) becomes indefinite when the velocity exceeds the Cherenkov threshold (more precisely when $|v| > c/(n - 1/n)$ with $n = \sqrt{\varepsilon\mu}$ the refractive index in the co-moving frame). In these conditions, the material system can become electromagnetic unstable and the wave energy may be negative. Two interacting material slabs with a relative velocity exceeding roughly two times the Cherenkov threshold may develop wave instabilities due to the spontaneous conversion of kinetic energy into light [31, 37-39]. These instabilities typically result in a quantum friction force [44] determined by the Abraham component of the Lorentz force, i.e., by Eq. (25) [31, 37].

**Acknowledgements:** This work was funded by Fundação para a Ciência e a Tecnologia under project PTDC/EEI-TEL/4543/2014 and by Instituto de Telecomunicações under project UID/EEA/50008/2013.